# MIMO Scheme Performance and Robust Detection in Mixture of Background and Impulse Noise

Sander Stepanov

Technion – Israel Institute of Technology

sanders@ee.technion.ac.il

## ABSTRACT

A new approach to analyzing and decoding MIMO signaling has been developed for the regular model of non-Gaussion noise, consisting of background and impulse noise named $\varepsilon$ - noise. It has been shown that non-Gaussion noise performance is significantly worse than Gaussion noise performance. Simulation results support the suggested theory. Robust in a statistical sense detection rule is suggested for such kind of noise, featuring a much better robust detector performance than the detector designed for Gaussian noise in an impulsive environment, but marginally less effective in background noise. The performance of the proposed approach is comparable with the developed potential bound. The proposed tool is a crucial issue for the MIMO communication system design, since a real noise environment has an impulsive character that contradicts the widely used Gaussian model. Therefore, the real MIMO performance for Gaussian and non-Gaussian noise models is entirely different.

## I. INTRODUCTION

Space-time coding has been a very popular issue in recent years. Most MIMO system designs and analyses have been developed for Gaussian noise, as for many other decoding schemes. The history of communication system designs provides many examples of using normal noise pdf first and using realistic non-Gaussian pdf afterwards. Recent publications [1-2] manifest that the MIMO theory gradually achieves this theoretically complicated level of non-Gaussion model statistical procedure designs. In [1], impulse noise (IN) means a background noise with a big dispersion, or only a Cauchy noise in a channel. However, IN actually exists together with background noise. In contrast to this research, in publication [2] the noise model used is limited by the assumption that the impulse noise realizations hit the whole slot or the whole epoch, but in reality, due to interleaving and space antenna diversity, noise realizations are entirely independent inside the same slot, i.e. noise parameters and realizations are independent in space and time.

The probabilistic model of additive noise should, on one hand, describe the fact that background noise is described by normal pdf and, on other hand, that there are moments of time when noise is increased considerably. The sources of IN are: industrial noise, frequency hopping and Ultra-Wide band interference, pulsed jamming, fast fading (in a sense that signal fades are amplified by after automatic gain control, then interleaved, finally ending up as IN) ; therefore, intensive investigations need to be carried  out to overcome the non-Gaussian character of noise [3-32, 46], suggesting the urgency of this research.

   The difficulties in analytical analysis are: the non-stationary nature of IN, and the complicated processing of IN in receiver circuits; however, joint existing background noise and IN at receiver antennas can be expressed by ε-noise added to sufficient statistic estimation [4, 6, 12, 13, 15, 21, 32].  In practice, accurate IN pdf estimation is difficult to accomplish, so some reasonable model of IN pdf should be presumed to adequately represent the random character of IN  (large entropy) and the significant power of IN. The normal pdf with a large variance can be used for this aim, since it reflects the fact that IN formatted by a large number of sources and big values of IN are expected [3, 4, 6, 8, 12,  13, 15, 21]. Along with pdf's, the normal pdf   for IN gives one the chance to calculate the decoding performance, as is shown in this paper. From the demonstrated examples of Alamouti's scheme (the performance of two transmiting antennas and two receiver antennas (2Tx2Rx)   [33] in Gaussian IN pdf and typical other IN pdf  - double exponent pdf [ 21 – 23, 32]) it follows that the normal model of IN pdf can be used as a first view estimation for other distributions as well. This research considers MIMO system performance using a mixture model of additive noise and an example of a typical MIMO scheme (Alamouti's scheme). Using this designated model provides a powerful analytical tool for MIMO system design, since the Gaussian noise point of support leads to non-effective statistical procedure design, when the real process model has a high incidence of significant noise realization. In MIMO practice, it can be helpful for an awareness of system performance in the early steps of a design and for the tailoring of analytical expectations and actual apparatus performance.

   As an accompanying result, the method for  MIMO scheme performance analysis in Gaussian noise was also developed, the example of a 2Tx2Rx scheme demonstrates its validity. The last publications in this field relate to 2-4-ary PSK with only one receiver antenna [34] (for other MIMO schemes, this approach is very complicated in use) or do not relate to actual MIMO schemes, but to an equivalent scaled AWGN channel [35].

   The analysis of the received equation for bit error rate (BER) for the traditional Minimum Least Square (MLS) decoder in IN leads to the conclusion that the decoder efficiency can be significantly improved by using robust in statistical sense theory  [37 – 40]. Proposed detection rule - The maximum likelihood of background noise residuals (MLBNR)  is a base for Robust Detector algorithms which achieve much better performance than the usual  MIMO decoding scheme in an impulsive environment, when deterioration  in Gaussian noise is minor.  The potential bound of  the MLBNR rule has been developed in this research in order to support further efforts to design more powerful Robust Detectors,  and to check the efficiency level of suggested algorithms.

The paper is organized as follows: section 2 introduces the system model; section 3 focuses on an analytical equation for Alamoti's scheme performance in non-Gaussian noise; in section 4 the results of calculations and simulations of Alamoui's scheme in $\varepsilon$-noise are discussed; section 5 consists of a description of the MLBNR rule and robust algorithm, algorithm performances are presented; and section 6 concludes the paper.  The reliability of simulation results was maintained in the following way: the simulations were run until 300 bit errors occurred.

## II. SYSTEM MODEL

The pdf of an additive noise mixture model comprising of background noise and independent impulse realizations can be described as [3,4,6,8,12,13,15,21]

$$\rho(\xi) = (1-\varepsilon)\varphi(\xi, \sigma_1^2) + \varepsilon\varphi(\xi, \sigma_2^2) \qquad (1)$$

where $\varphi(\xi, \sigma^2)$ is the Gaussian pdf with variance $\sigma^2$ and zero means.

When $\xi$ is complex, it means that the real and the imaginary part have the same variance and PDF. This model is an approximation of the more general Middleton Class A noise model [9, 13]. According to Alamouti MIMO scheme [33], these are the signals in received antennas

$$\begin{aligned} r_1 &= r_{11} + r_{21} \\ r_2 &= r_{12} + r_{22} \end{aligned} \qquad (2)$$

where

$r_1$ and $r_2$ - efficient statistic estimations at decoder input from the first and second antennas;

$$\begin{aligned} r_{11} &= S_1 h_{11} + (-S_2) h_{21} + \xi_{11} \\ r_{21} &= S_1 h_{12} + (-S_2) h_{21} + \xi_{21} \\ r_{12} &= S_2 h_{11} + S_1 h_{21} + \xi_{12} \\ r_{22} &= S_2 h_{12} + S_1 h_{22} + \xi_{22} \end{aligned}$$

$r_{ji}$ and $\xi_{ji}$ - the signal and noise efficient statistic estimation from antenna $j$ and at epoch $i$;

$S_1$ and $S_2$ – transmitted symbols at slot .

Let us note here that noise above can be completely, independent in the sense of equal probability of pdf variance, or here can be a functional connection between variances: noise realizations at the same epoch have the same variance, and noise realizations at the same slot have the same variance. A more complicated, completely independent case is analyzed here, from which other cases can be deduced.

## III. PERFORMANCE ANALYSIS FOR THE MLS DECISION.

The analysis will be carried out by an investigation of the probability of error for a definite pair ($S_1$, $S_2$) and its alternative pair ($S'_1$, $S'_2$) for 2Tx2Rx Alomouti's scheme. The other more complicated MIMO scheme cases can be analyzed by analogue. The efficient statistic estimations of signals only (without noise) are

$$\begin{aligned} x_{11} &= S_1 h_{11} + (-S_2) h_{21} \\ x_{21} &= S_1 h_{12} + (-S_2) h_{21} \\ x_{12} &= S_2 h_{11} + S_1 h_{21} \\ x_{22} &= S_2 h_{12} + S_1 h_{22} \end{aligned}$$

and

$$x_{11} = S'_1 h_{11} + (-S'_2) h_{21}$$
$$x_{21} = S'_1 h_{12} + (-S'_2) h_{21}$$
$$x_{12} = S'_2 h_{11} + S'_1 h_{21}$$
$$x_{22} = S'_2 h_{12} + S'_1 h_{22}$$

The probability of error in detection in $(S'_1, S'_2)$ instead of $(S_1, S_2)$ is

$$P_{se} = P\left(\begin{array}{l}[(r_{11} - x_{11})^2 + (r_{21} - x_{21})^2 + (r_{12} - x_{12})^2 + (r_{22} + x_{22})^2] > \\ [(r_{11} - x'_{11})^2 + (r_{21} - x'_{21})^2 + (r_{12} - x'_{12})^2 + (r_{22} + x'_{22})^2]\end{array}\right) \quad (3)$$

$$P_{se} = P[|\xi_{11}|^2 + |\xi_{21}|^2 + |\xi_{12}|^2 + |\xi_{21}|^2 >$$
$$|(S_1 h_{11} + (-S_2) h_{21} + \xi_{11} - S'_1 h_{11} - (-S'_2) h_{21})|^2 +$$
$$|(S_1 h_{12} + (-S_2) h_{21} + \xi_{22} - S'_1 h_{12} + (-S'_2) h_{22})|^2 + \quad (4)$$
$$|(S_2 h_{11} + S_1 h_{21} + \xi_{12} - S'_2 h_{11} + S'_1 h_{21})|^2 +$$
$$|(S_2 h_{12} + S_1 h_{22} + \xi_{22} - S'_2 h_{12} + S'_1 h_{22})|^2]$$

Designate the differences as $e_1 = S_1 - S'_1$ and $e_2 = S_2 - S'_2$. The upper bound of $P_{se}$ can be expressed by using the error of detection for the minimum Euclid distance pair, when $e_2=0$. Later, the $\hat{P}_{se}$ stands for the upper bound of error in this sense:

$$\hat{P}_{se} = P[|\xi_{11}|^2 + |\xi_{21}|^2 + |\xi_{12}|^2 + |\xi_{21}|^2 > |(e_1 h_{11} + \xi_{11})|^2 + |(e_1 h_{12} + \xi_{21})|^2 +$$
$$|(e_1 h_{21} + \xi_{12})|^2 + |(e_1 h_{22} + \xi_{22})|^2], \quad (5)$$

this can be simplified to

$$\hat{P}_{se} = P[|h_{11}|\xi_1 + |h_{12}|\xi_2 + h_{21}|\xi_3| + |h_{22}|\xi_4 > d_{min}^2]. \quad (6)$$

where $\xi_i$ is still distributed by (1), but is real (in contrast to equations (4)-(5)).

$$d_{min}^2 = e_1(|h_{11}|^2 + |h_{21}|^2 + |h_{12}|^2 + |h_{21}|^2).$$

Now by using Appendix I, $\hat{P}_{se}$ can be expressed as

$$\hat{P}_{se} = \sum_{i_1=0}^{i_1=1} \cdots \sum_{i_4=0}^{i_4=1} \left(\prod_{n=1}^{n=4}(1-\varepsilon)(1-i_n) + \varepsilon i_n\right) Q\left(\frac{d_{min}^2/2}{\sqrt{\begin{array}{l}|h_{11}|^2((1-i_1)\sigma_1^2 + i_1\sigma_2^2) + |h_{12}|^2((1-i_2)\sigma_1^2 + i_2\sigma_2^2) + \\ |h_{21}|^2((1-i_3)\sigma_1^2 + i_3\sigma_2^2) + |h_{22}|^2((1-i_4)\sigma_1^2 + i_4\sigma_2^2)\end{array}}}\right), \quad (7)$$

where

$$Q(x) = \frac{1}{\sqrt{2\pi}} \int_x^\infty e^{-t^2/2} dt.$$

By averaging all possible sets of channel coefficients and possible impulse noise scenarios, the bit general error probability $P_e$ is given by

$$P_e = \sum_{i_1=0}^{i_1=1} \bullet \bullet \bullet \sum_{i_4=0}^{i_4=1} \left( \prod_{n=1}^{n=4} (1-\varepsilon)(1-i_n) + \varepsilon i_n \right) \int_{-\infty}^\infty \int_{-\infty}^\infty \int_{-\infty}^\infty \int_{-\infty}^\infty \varphi(h_{11}, 0.5)\varphi(h_{12}, 0.5)\varphi(h_{21}, 0.5)\varphi(h_{22}, 0.5) \hat{P}_{se} dh_{11} dh_{12} dh_{21} dh_{22}.$$

(8.1)

(note the error in the minimum Euclid distance pair leads to a one bit error, and the other pair error leads to a 2 bit error, therefore this $P_e$ is not the upper bound but a tight approach)

BER performance in Gaussian noise can be used for (8.1) evaluation in the following way: note that the BER for only the Gaussian noise case $P_{beG}$ can be expressed only by integrals part of (8.1) if $\hat{P}_{se}$ were replaced by the probability of error between pairs $(S_1, S_2)$ and $(S'_1, S'_2)$ in Gaussian noise ($P_{seG}$)

$$P_{beG} = \int_{-\infty}^\infty \int_{-\infty}^\infty \int_{-\infty}^\infty \int_{-\infty}^\infty \varphi(h_{11}, 0.5)\varphi(h_{12}, 0.5)\varphi(h_{21}, 0.5)\varphi(h_{22}, 0.5) P_{seG} dh_{11} dh_{12} dh_{21} dh_{22}. \quad (8.2)$$

This $P_{seG}$ depends only on one parameter noise variance $\sigma^2$ for given channel coefficients, so if $\hat{P}_{se}$ were replaced by $P'_{se}$, which is

$$\hat{P}_{se} \approx P'_{se} = Q\left( \frac{d_{\min}^2 / 2}{\sqrt{(h_{11}^2 + h_{12}^2 + h_{21}^2 + h_{22}^2) \sum_{k=1}^{k=4} ((1-i_k)\sigma_1^2 + i_k \sigma_2^2)/4}} \right) \quad (9)$$

then

$$P_{be} \approx \sum_{i_1=0}^{i_1=1} \bullet \bullet \bullet \sum_{i_4=0}^{i_4=1} \left( \prod_{n=1}^{n=4} (1-\varepsilon)(1-i_n) + \varepsilon i_n \right)$$

$$\int_{-\infty}^\infty \int_{-\infty}^\infty \int_{-\infty}^\infty \int_{-\infty}^\infty \varphi(h_{11}, 0.5)\varphi(h_{12}, 0.5)\varphi(h_{21}, 0.5)\varphi(h_{22}, 0.5) P_{seG}(\sum_{k=1}^{k=4} ((1-i_k)\sigma_1^2 + i_k \sigma_2^2)/4) dh_{11} dh_{12} dh_{21} dh_{22} =$$

$$= \sum_{i_1=0}^{i_1=1} \bullet \bullet \bullet \sum_{i_4=0}^{i_4=1} \left( \prod_{n=1}^{n=4} (1-\varepsilon)(1-i_n) + \varepsilon i_n \right) P_{beG}(\sum_{k=1}^{k=4} ((1-i_k)\sigma_1^2 + i_k \sigma_2^2)/4)$$

(10.1)

or in simplified form

$$P_{be} = (1-\varepsilon)^4 P(\sigma_1^2) + 4(1-\varepsilon)^3 \varepsilon P(\frac{3\sigma_1^2 + \sigma_2^2}{4}) + 6(1-\varepsilon)^2 \varepsilon^2 P(\frac{\sigma_1^2 + \sigma_2^2}{2}) +$$
$$4(1-\varepsilon) \varepsilon^3 P(\frac{\sigma_1^2 + 3\sigma_2^2}{4}) + \varepsilon^4 P(\sigma_2^2)$$
(10.2)

where

$P(\sigma^2)$ - is the probability error for only Gaussian case noise with a variance of $\sigma^2$.

For a current 2Tx2Rx example, $P(\sigma^2)$ can be taken from [33] simulation results or in a general case calculated by the approach developed in Appendix II.

## IV. SIMULATION RESULTS FOR 2TX2RX MLS DECODER

The performances of 2Ttx2Rx MLS decoder for typical values $\gamma$, which range from 10 to 100 [21] and for rarely occurring ($\varepsilon = 0.01$) and for frequently occurring ($\varepsilon = 0.1$) impulse noise realizations, are depictured Fig.1 to Fig. 6 (here demonstrated results are not only for end range values but also for intermediate $\varepsilon$ value 0.05 and $\gamma$ value 33 and for some outside the typical $\gamma$ range, to show the performance "saturation" dynamics as well) . The cases when an impulsive noise tail has a Double exponent (Laplas) pdf [ 6, 18 – 20, 32, 37]

$$\rho(\xi) = (1-\varepsilon)\varphi(\xi,\sigma_1^2) + \varepsilon(2\sigma_2^2)^{-1/2} e^{-|\xi| (2/\sigma_2^2)^{1/2}}$$
(11)

and when whole noise pdf is a Double exponent are also represented in these figures.

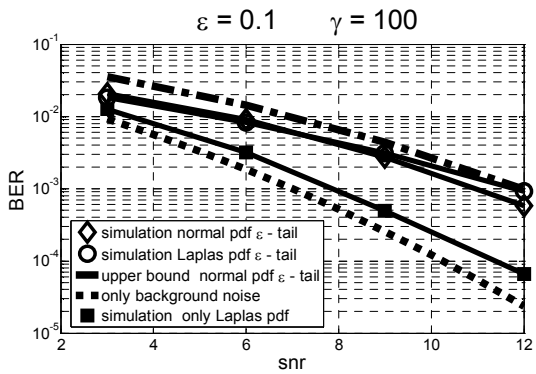
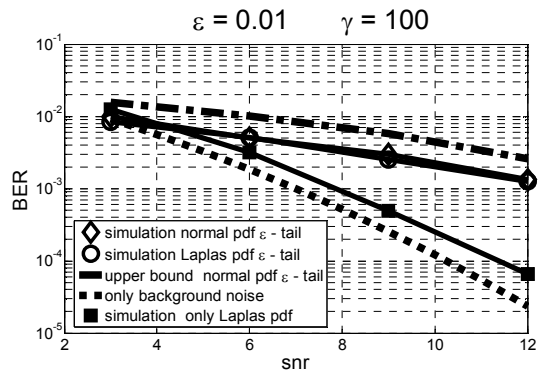

Fig 1                     Fig2

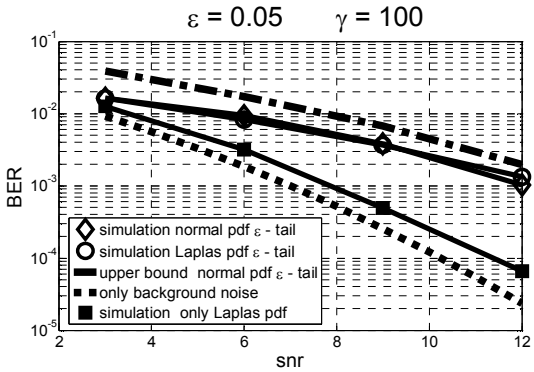

Fig 3

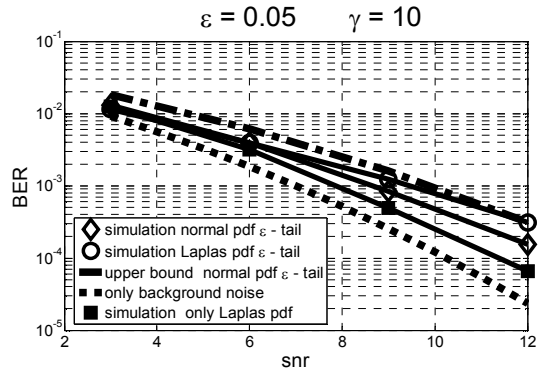

Fig 4

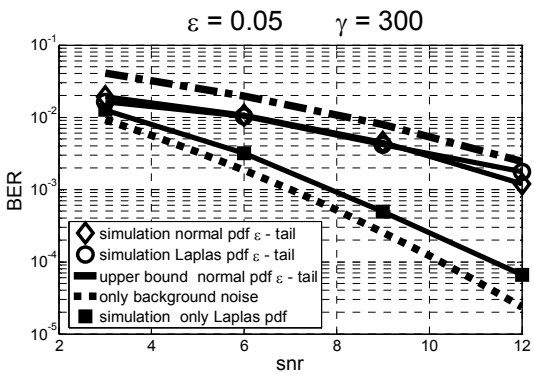

Fig 5

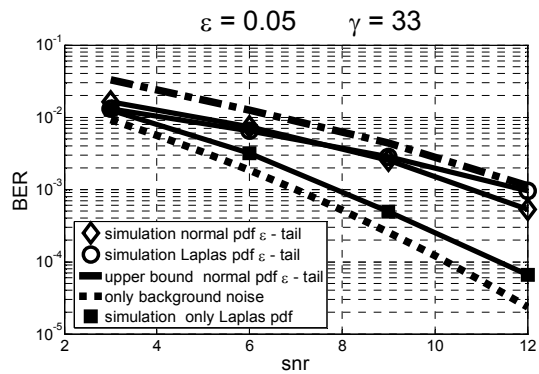

Fig 6

Fig.1 – Fig.6 the analytical and simulation performances of 2Ttx2Rx MLS decoder for typical values $\gamma$ and $\varepsilon$, when the impulsive noise pdf is Gaussian and Double exponent. Simulation performance when whole noise distribution is Laplas.

To explore the behavior of nonlinear performance dynamics in a wide area of possible noise parameters situation, the simulation was carried out using the $\varepsilon = 0.05$  $\gamma = 100$ set of parameters as a starting point and afterwards the $\varepsilon$ was increased to 0.1 and reduced to 0.01, and $\gamma$ was reduced by two steps in each 3 times, and increased by one step 3 times.

As is shown in Fig.1-Fig.6, $\varepsilon$-noise dramatically influences the decoder performance. The main points are:

1. Close inspection of Fig.1- Fig.3 shows that if the probability of IN falls from "often" to "rare" when its power is big, BER it gradually returns to only a Gaussian noise case. For practical use, it means that it does not matter if the impulse noise is "often" or "rare", it leads to nearly the same significant performance deterioration, when the intensity of IN is high.

2. Referring to Fig.3, Fig.5 and Fig.6, it is observed that for big $\gamma$ the BER does not depend on the value of $\gamma$. It can be explained by the following: on one hand - $\sigma_2^2 \approx \sigma^2/\varepsilon$, on the other hand, when the impulse noise realizations hit the slot it is like low snr detection conditions, but when snr is low, the BER curve slopes gradually, therefore BER varies slightly according to the power of the impulse noise. The conclusion is: if strong $\varepsilon$-noise realization causes error, the enlargement of $\varepsilon$-noise accomplishes nothing.

3. Even non-significant $\varepsilon$-noise conditions differ from only a Double exponent pdf case, Fig.4.

4. Suggested upper bound serves BER estimation well, in a wide range of noise parameters.

5. Gaussian $\varepsilon$-noise provides a good choice of $\varepsilon$-noise pdf, since heaver "tails" - the double-exponent $\varepsilon$-noise - cause nearly the same BER performance. The inference is: according to demonstrated examples of the Alamouti scheme performance in double - exponent distribution as a "tail" of $\varepsilon$-noise, it follows that a normal model pdf of IN can be used as a first view estimation for other distributions as well.

It can be useful to compare quantitatively the addends from equation (10. 2) with BER: therefore, the values of BER and corresponding addends are shown in Fig.7 and Fig. 8.

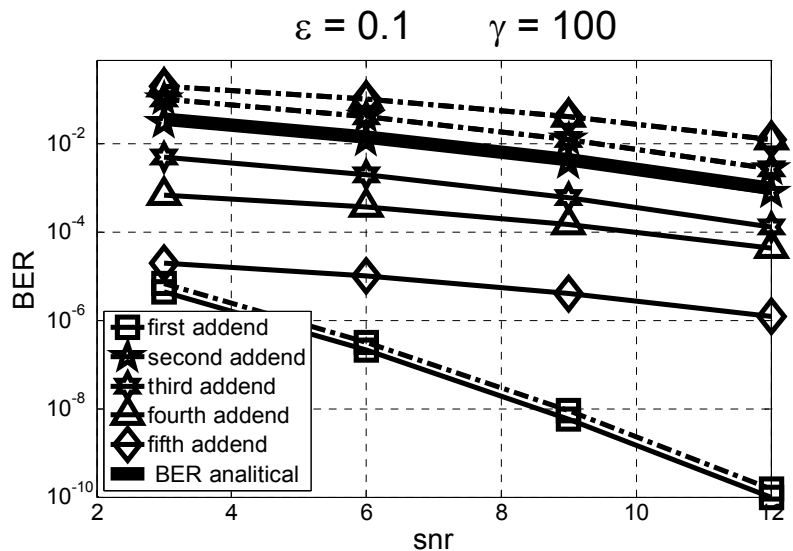

Fig.7

Fig. 7 BER and addends in the equation (10. 2) for MLS detector in "often" impulsive $\varepsilon$-noise; solid lines illustrate the addends themselves, and dashed lines only the P(.) value from addends.

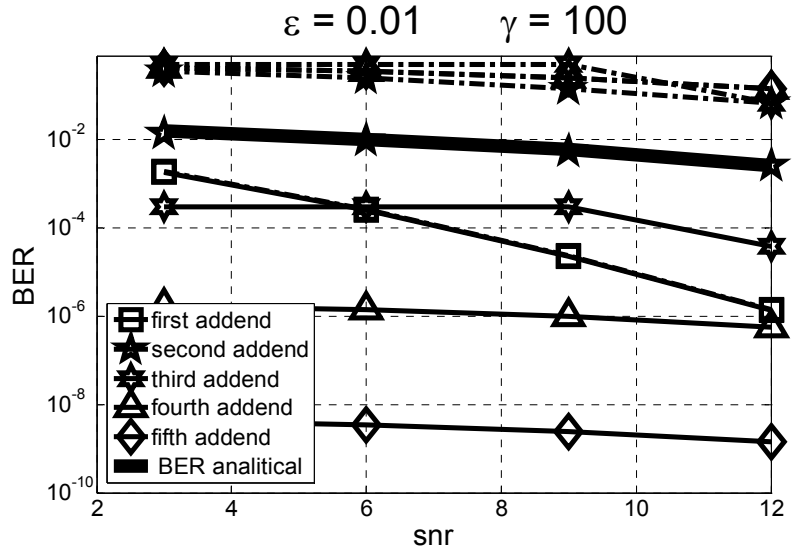

Fig. 8

Fig. 8 BER and addends in the equation (10. 2) for MLS detector in "rare" impulsive $\varepsilon$-noise, solid lines illustrate the addends themselves, and dashed lines only the P(.) value from addends.

From these figures it follows that the first addend contributes a very small part to the whole BER, since the background noise power is relatively small compared to the IN one, and that the main contribution is from the second addend. Note, the P(.) values of the other addends can also be big, but their coefficients at P(.) are very small, therefore they do not play the main role. This note leads to an important conclusion: it is worthwhile making the decision based on background noise realizations, then BER can be reduced toward the first addend values.

## V. MIMO ROBUST DETECTORS

From the small values of addend number 1 in (10), an important heuristic conclusion follows: it is worthwhile making the decision only based on residuals formatted by background noise (analytical justification of this inference can be found in Appendix III).  Such a decision is the rule for the minimum probability of vector detection error (such a criterion, defined for only vector signals brings small bit probability errors as well, as can be seen from Algoritm Viterbi's wide use),  as follows:

$$\hat{S} = \arg\min_{S} \sum_{\substack{j=1 \\ i,j \in Y}}^{j=m} \sum_{i=1}^{i=n} |r_{ij} - x_{ij}|^2, \qquad (12)$$

where    $S$ - all possible signal vectors to be transmitted;

$\hat{S}$ - the decision about transmitted signals;

$n$ – the number of Tx antennas;

$m$ - the number of Rx antennas;

$Y$ – the set of $\xi_{i,j}$ formatted only by background noise.

which is the Maximum Likelihood of Background Noise Residuals (MLBNR).

Of course this MLBNR rule can not be exactly accomplished in practice, as it is impossible to know those residuals formatted by background noise, but this is a technique whereby a simple practical approach can be deduced by taking into account that the equation (12) can be rewritten in the following form:

$$\hat{S} = \arg\min_{S} \sum_{\substack{j=1 \\ i,j \in Y}}^{j=m} \sum_{i=1}^{i=n} \left|\hat{\xi}_{ij}\right|^2 \qquad (13)$$

where $\sum_{\substack{j=1 \\ i,j \in Y}}^{j=m} \sum_{i=1}^{i=n} \left|\hat{\xi}_{ij}\right|^2$ resembles the scale parameter estimate in Robust Statistics terms [37 - 40] in the method of calculation, therefore the strong and common practiced procedures of Robust Statistics can be used to improve the BER in $\varepsilon$-noise.

The simplest Robust detector can be based on the median estimate of residuals (M-decoder), according to

$$\hat{S} = \arg\min_{S} median\left(\left|\hat{\xi}_{ij}\right|\right) \qquad (14)$$

where $i = 1,...,n; j = 1,...,m$

The aim of this paper is not to carry out a survey of all possible Robust Estimators from the point of view of finding the best Robust Decoder, this would be an issue for another project. However, what is possible is to show the performance relation between a simple M-decoder and a more complicated Robust Decoder. At this point, let the W-decoder be introduced, which uses the more complicated but more effective W-estimate for further BER reduction. The decision rule is (15) where coefficients $w_{ij}$ are computed interactively by starting from point $w_{ij} = 1$ for $\forall_{IJ}$. Further $w_{ij}$ is an update for each iteration by equations (16)

$$\hat{S} = \arg\min_{S} \frac{\sum_{j=1}^{j=m}\sum_{i=1}^{i=n} w_{ij}\left|\hat{\xi}_{ij}\right|}{\sum_{j=1}^{j=2}\sum_{i=1}^{i=2} w_{ij}} \qquad (15)$$

$$w_{ij} = (1-u_{ij}^2)^2 \quad \text{if} \quad |u_{ij}|>1$$

$$w_{ij} = 0 \quad \text{if} \quad |u_{ij}|<1 \qquad (16)$$

where

$$u_{ij} = \frac{\left|\hat{\xi}_{ij}\right| - M}{\left(\min(\left|\hat{\xi}_{ij}\right|) + \max\frac{(\left|\hat{\xi}_{ij}\right|)}{2}\right) P_2} \qquad (17)$$

$$M = \frac{\sum_{j=1}^{j=m}\sum_{i=1}^{i=n} w_{ij}|\hat{\xi}_{ij}|}{\sum_{j=1}^{j=m}\sum_{i=1}^{i=n} w_{ij}} \qquad (18)$$

$P_2$ -robustness regulation parameter

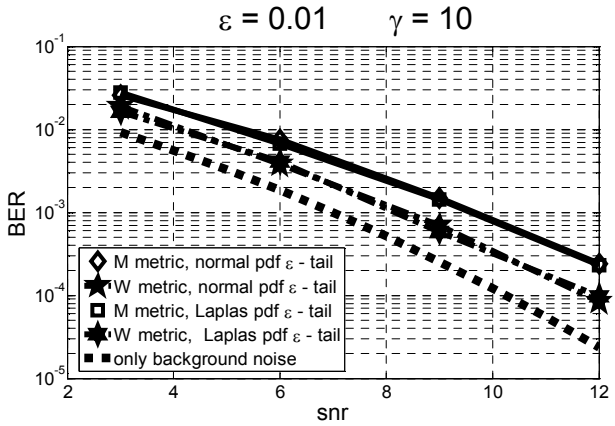

Fig. 9

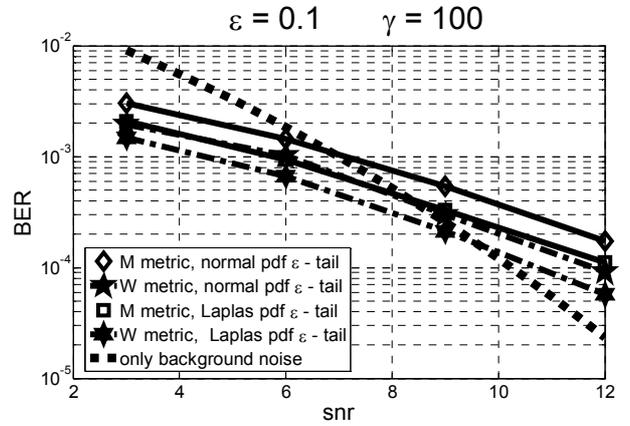

Fig. 10

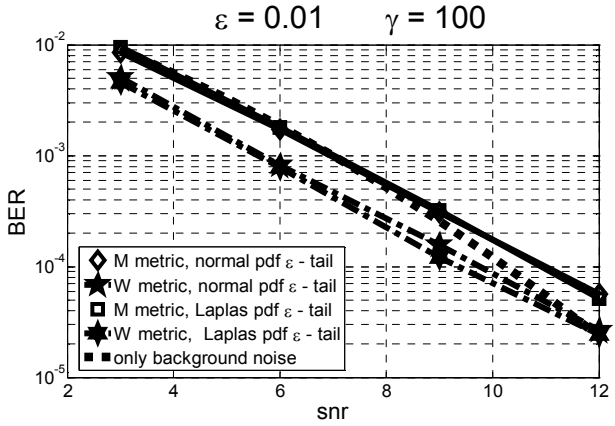 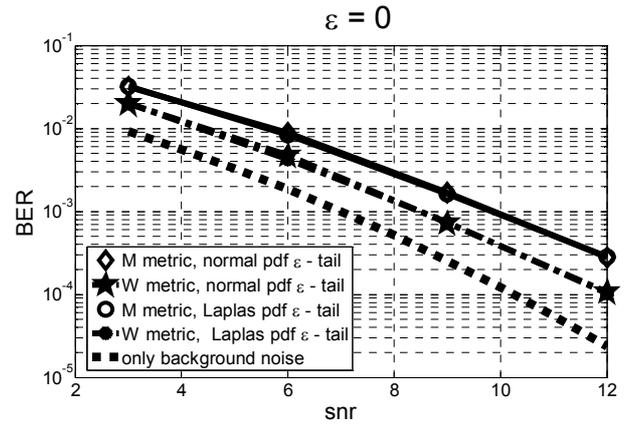

Fig. 11                                    Fig. 12

Fig.9-Fig.12 The analytical performance 2Tx2Rx of Robust Decoders.

It is possible to see from Fig.9-Fig.12 that a Robust M-decoder and a W-decoder produce better performances than a classical MLS MIMO decoder-the same as for normal ε-tail for Laplas ε-tail. Note that Robust Decoder performance curves are close to BER curves for the MLS decoder is only background noise, which in turn goes much lower than the corresponding curves for the MLS decoder in ε-noise: see Fig.1-Fig.6.

The M-decoder is much simpler than the W-decoder, nevertheless, its performance is only nearly 1 dB worse. The difference between a Laplas-tail performance and a normal one is small, which can be expected since Robust statistics are designed to be invariant to tail distribution. The gain from the use of Robust Decoders is close to the loss MLS of the decoder performance in IN, compared with the MLS decoder in background noise. In other words, the Robust decoder returns the decoder performance to background noise performance when the noise is impulsive, i.e. the Robust Decoder is invariant to noise pdf "tails".

In order to make it possible to compare the Robust Decoder performance with a potentially achievable one (normal ε-tail), the BER of optimal in a MLBNR sense decoder can be calculated by averaging by probability the probability of errors of the MLS decision by relay on background noise samples or else on normal ε-tail samples, if only they are. For Appendix II a 2Tx2R example, the MLBNR optimal decoder BER is

$$P_{eRobust} = \sum_{i=0}^{3} C_i^4 (1-\varepsilon)^{4-i} \varepsilon^i \int_0^\infty 2Chi(2x, 8-2i) Q\left(\frac{\sqrt{x}}{\sigma_1 \sqrt{2}}\right) dx + \varepsilon^4 \int_0^\infty 2Chi(2x, 8) Q\left(\frac{\sqrt{x}}{\sigma_2 \sqrt{2}}\right) dx . \qquad (19)$$

where $C_0^4 = 1$

The last addend in (19) means that, if there are no background samples, it is still better to make the decision to use IN noise samples, than to do nothing and choose some value of bit to be transmitted, with probability 0,5 to make the right choice. Fig. 13 and Fig. 14 represent the values of addends from (19) together with BER received by simulation for W-decoder from Fig 10-Fig 11 and BER for optimal MLBNR detector from (19).

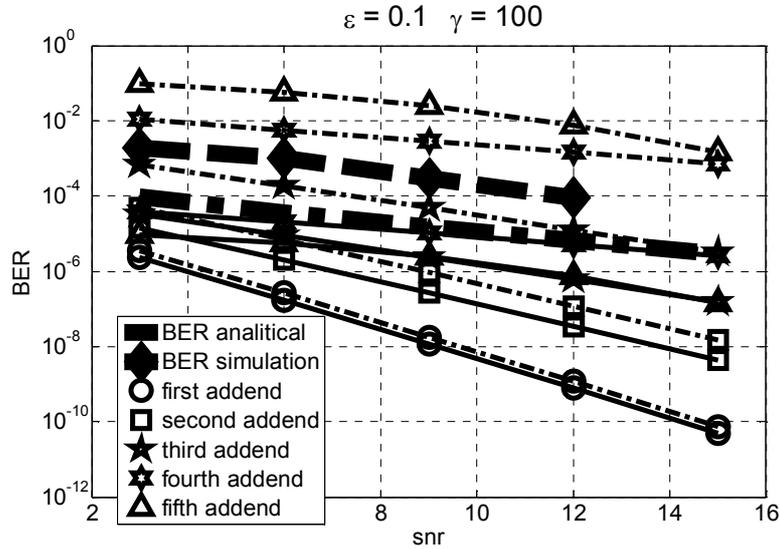

Fig. 13

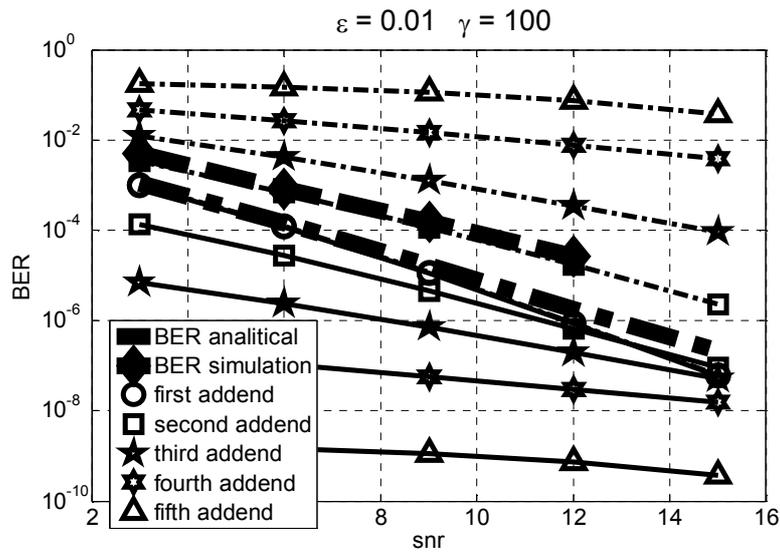

Fig. 14

Fig.13 - Fig.14 Analytical MLBNR BER in "often" and "rare" IN for 2Tx2Rx example and its addends (wide dashed line), these addends before averaging by probability (dashed lines) , BER received by simulation for W-decoder (wide diamond line) .

It can be seen that the W-decoder achieves good performances in comparison with a potentially possible performance. The dashed line represents BER for each possible combination of background and IN samples before averaging by probability. The dashed lines give the example of quantitative analysis of a situation which would be possible if the number of antennas were reduced (addends 1-4 from (19)) or what would happen if all antennas were hit by impulse noise (last addend in (19)). The interpretation is: in "often" IN should be strengthen the decoding in presence of IN samples (first addend much lower than achievable found); in "rare" IN should be strengthen the decoding in only background noise samples case (first addend higher than other addends).

## APPENDIX I

For convenience let us designate

$h_1 = |h_{11}|, h_2 = |h_{21}|; h_3 = |h_{12}|; h_4 = |h_{21}|$

and treat the channel coefficients as constant. For calculation of (6) we need the equation for pdf of the random variable

$$\eta = |h_{11}|\xi_1 + |h_{12}|\xi_2 + |h_{21}|\xi_3| + |h_{22}|\xi_4 = \sum_{i=1}^{4} \xi_i |h_i|, \quad (AI.1)$$

then the characteristic function of $\eta$ is

$$\Theta(z) = \prod_{i=1}^{4} \Theta_i(z), \quad (AI.2)$$

where

$\Theta_i(r)$ - is a characteristic function of random variable $\xi_i |h_i|$

$$\Theta_i(z) = \int_{-\infty}^{\infty} \rho(\eta_i) \exp(jz\eta_i) d\eta_i$$

from (1) follows

$$\Theta_i(z) = \int_{-\infty}^{\infty} \left[ (1-\varepsilon)\varphi(\eta_i, 0, (\sigma_1 |h_i|)^2) + \varepsilon\varphi(\eta_i, 0, (\sigma_2 |h_i|)^2) \right] \exp(jz\eta_i) d\eta_i = (1-\varepsilon)\Theta_{1i}(z) + \varepsilon\Theta_{2i}(z) \quad (AI.3)$$

where

$$\Theta_{1i}(z) = \int_{-\infty}^{\infty} \varphi(\eta_i, 0, (\sigma_1 |h_i|)^2) \exp(jz\eta_i) d\eta_i,$$

and

$$\Theta_{2i}(z) = \int_{-\infty}^{\infty} \varphi(\eta_i, 0, (\sigma_2 |h_i|)^2 (\exp(jz\eta_i) d\eta.$$

Then

$$\Theta(z) = \prod_{i=1}^{4}[(1-\varepsilon)\Theta_{1i}(z) + \varepsilon\Theta_{2i}(z)] = \sum_{i_4=0}^{i_4=1}...\sum_{i_1=0}^{i_1=1}[(1-i_1)(1-\varepsilon)\Theta_{11}(z) + i_1\varepsilon\Theta_{21}(z)] *$$

$$*[(1-i_2)(1-\varepsilon)\Theta12(z) + i_2\varepsilon\Theta_{22}(z)]...[(1-i_4)(1-\varepsilon)\Theta_{14}(z) + i_4\varepsilon\Theta_{24}(z)]$$

(AI. 4)

The needed pdf can be expressed from (AI. 4), therefore the probability (6) is

$$P_{se} = \sum_{i_4=0}^{i_4=1}...\sum_{i_1=0}^{i_1=1} 1/2\pi \int_{d^2_{min/2}}^{\infty}\int_{-\infty}^{\infty}[(1-i_1)(1-\varepsilon)\Theta_{11}(z) + i_1\varepsilon\Theta_{21}(z)][(1-i_2)(1-\varepsilon)\Theta_{12}(z) + i_2\varepsilon\Theta_{22}(z)]$$

$$...[(1-i_4)(1-\varepsilon)\Theta_{14}(z) + i_4\varepsilon\Theta_{24}(z)]\exp(-jz\eta)dzd\eta$$

(AI. 5)

for the inner integral of (AI. 5)

$$1/2\pi\int_{-\infty}^{\infty}[(1-i_1)(1-\varepsilon)\Theta_{11}(z) + i_1\varepsilon\Theta_{21}(z)]...[(1-i_4)(1-\varepsilon)\Theta_{14}(z) + i_4\varepsilon\Theta_{24}]\exp(-jx\eta)dz = [(1-i_1)(1-\varepsilon) + i_1\varepsilon]$$

$$...[(1-i_4)(1-\varepsilon) + i_4\varepsilon]\psi(\eta)$$

(AI. 6)

where ψ(η) is the pdf of

$$\eta = (1-i_1)\eta_{11} + i_1\eta_{21} + ... + (1-i_4)\eta_{14} + i_4\eta_{24};$$

with the pdf of $\eta_{1i}$

$$\varphi(\eta_{1i}, 0, (\sigma_1|h_i|)^2);$$

and the pdf of $\eta_{2i}$

$$\varphi(\eta_{2i}, 0, (\sigma_2|h_i|)^2).$$

From (AI. 5) and (AI. 6) follows equation (7).

## APPENDIX II

In this Appendix the approach is to calculate BER in background noise for arbitrary MIMO schemes. The detector operates with vector signals, therefore according to [34, 41] the tight upper bound of BER can be calculated by using the probability of deciding erroneously between two vector signals with minimum Euclid distance between them, and taking into account the number of bit errors between these vectors.

Mathematically, in a general case (the equal power of transmitted symbols and the arbitrary number of transmission and receiving antennas) in order to estimate the vector error rate $P_{eV}$, it is necessary to evaluate the probability to be true to the equation

$$P\left(\xi > \frac{e\sqrt{d^2_{min}}}{2\sigma}\right) \qquad (AII.1)$$

where

$\xi$ – represents the Gaussian noise;

$\sigma$ – the variance of Gaussian noise;

$d^2_{min}$ – the variable distributed according to $2Chi(2x,n)$

where $Chi(x,n)$ is the pdf of Chi-squared distribution with the degree n

$$Chi(x,n) = \frac{1}{2^{\frac{n}{2}}\Gamma(\frac{n}{2})} x^{(\frac{n}{2}-1)} e^{\frac{-x}{2}}$$

and

$$\Gamma(x) = \int_0^\infty t^{(x-1)} e^{-t} dt$$

for real x and

$$\Gamma(n+1) = n!$$

for integer x, $e$ – the absolute value of the difference between transmitted symbols.

Here, namely the pdf $2Chi(2x,n)$ is used, since due to normalization used in [34] the channel coefficient in each dimension has a variance of ½.

Now $P_{eV}$ can be expressed by averaging by the probability of variable $d^2_{min}$

$$P_{eV} = \int_0^\infty 2Chi(2x,n) Q\left(\frac{\sqrt{x}}{\sigma\sqrt{2}}\right) dx \qquad (AII.2)$$

For floating point n, the needed integral can be expressed as

$$P_{eV} = 2K \left( \frac{2^{(-2+\frac{1}{2}n)}\Gamma(\frac{1}{2}n) - 2^{(-2+\frac{1}{2}n)} {}_2F_1([\frac{1}{2},\frac{1}{2}+\frac{1}{2}n],[\frac{3}{2}],-\frac{1}{4}\frac{1}{\sigma^2})\Gamma(\frac{1}{2}+\frac{1}{2}n)}{\sigma\sqrt{\pi}} \right), \qquad (AII.3)$$

where

$$K = \frac{1}{2^{n/2}\Gamma(n/2)},$$

${}_2F_1([a,b],c,z)$ is a Gauss hypergeometric function

$$_2F_1([a,b],c,z) = \frac{\Gamma(c)}{\Gamma(b)\Gamma(c-b)} \int_0^1 \frac{t^{b-1}(1-t)^{c-b-1}}{(1-tz)^a} dt.$$

This integral taken by using the table integral and by changing the integral variable to $y = x^2$ [42]

$$\int_0^\infty x^{\alpha-1} e^{-px^2} erf(cx) dx = \frac{c}{\sqrt{\pi} p^{(\alpha+1)/2}} \Gamma\left(\frac{\alpha+1}{2}\right) {}_2F_1\left(\frac{1}{2}, \frac{\alpha+1}{2}; \frac{3}{2}; -\frac{c^2}{p}\right)$$

The upper bound simplification can be achieved by using the next approximation

$$Q(x) < \hat{Q}(x) = \frac{1}{x\sqrt{2\pi}} e^{\frac{-x^2}{2}} \qquad (AII.\ 4)$$

Then $P_{eV}$ can be approximated by

$$\hat{P}_{eV} = \int_0^\infty \frac{2 Chi(x2, n) e^{\frac{-x}{4\sigma^2}}}{\sqrt{2\pi x}} dx, \qquad (AII.\ 5)$$

$$\sigma\sqrt{2}$$

taking this integral by the table integral from [45]

$$\int_0^\infty x^{\upsilon-1} e^{-\mu x} = \frac{1}{\mu^\upsilon} \Gamma(\upsilon)$$

we have

$$\hat{P}_{eV} = \frac{\sigma(2\sigma)^{(n-1)}}{\sqrt{\pi}(1+4\sigma^2)^{(\frac{n}{2}-\frac{1}{2})}} \frac{\Gamma(\frac{n}{2} - \frac{1}{2})}{\Gamma(\frac{n}{2})}. \qquad (AII.\ 6)$$

For the definite Alamouti scheme used in this work the BER estimation and its upper bound are

$$\hat{P}_{eG} = \frac{1}{2} - \frac{35}{64} \frac{{}_2F_1([\frac{1}{2}, \frac{9}{2}], [\frac{3}{2}], -\frac{1}{4}\frac{1}{\sigma^2})}{\sigma} \qquad (AII.\ 7)$$

and

$$\hat{P}_{eG} = \frac{70.89 * \sigma^8}{\sqrt{\pi}(1+4\sigma^2)^{(3.5)}}. \qquad (AII.\ 8)$$

From Fig. AII. 1 it can be seen that the developed equations for BER estimation are tight to simulation curve results from [33].

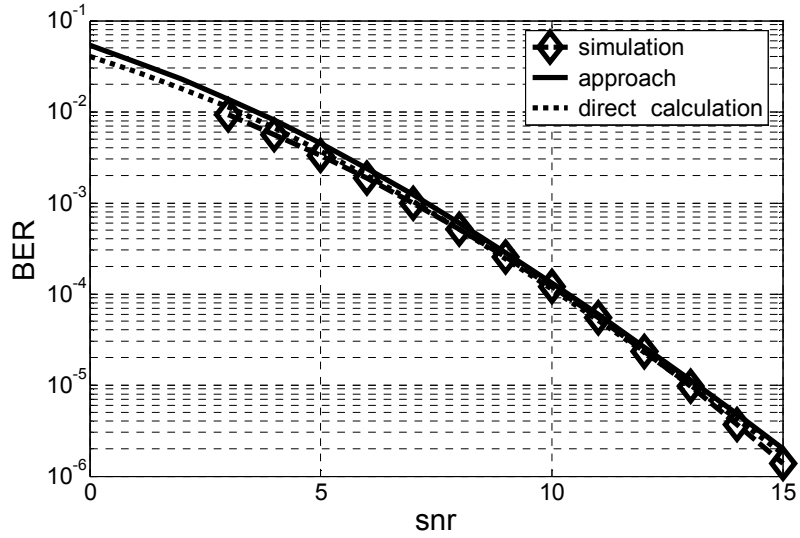

Fig. AII. 1

Fig. AII. 1 2Tx2Rx scheme performance in background noise for simulation, approach (AII. 8) and direct calculation (AII. 7)

Note that in [35] the calculation of integral AII. 2 for integer n was described. In a general case it can be necessary to use floating point n for complicated MIMO schemes. For example, if one more antenna were added to the transmitter having transmission $-S_1$ at the first epoch and $-S_2$ at the second epoch with channel coefficients $h_{31}$ and $h_{32}$ towards the first and second receiver antennas, then the probability of a mistaken decision in favor of the transmitted pair 1 and -1, instead of the transmitted 1 and 1, can be expressed by analogue with the previous 2Tx2Rx case, as

$$P_{ev} = Q\left(\frac{e\sqrt{|h_{21}|^2 + |h_{22}|^2 + |h_{11} - h_{31}|^2 + |h_{12} - h_{32}|^2}}{\sigma}\right). \qquad \text{(AII. 9)}$$

The addends inside the root equation have not the same variance, since the inside root sum is not ruled by Chi-squared distribution. The situation can be overcome by a traditional approximation for unknown pdfs by a Gaussian pdf with mean $M_A$ and variance $\sigma_A^2$. Those parameters can be calculated since the distributions for the addends are known [44]. This Gaussian pdf, in its turn, can be approximated by a Chi-squared distribution equation [45], in the following way. If the random variable is distributed according to a Chi-squared pdf with the degree n - $\chi_n$ is divided by parameter *a*, then the pdf of variable $\chi_n/a$ is $a*Chi(ax,n)$, with mean $M(\chi_n/a) = n/a = M_A$; and variance $\sigma^2(\chi_n/a) = 2n/a^2 = \sigma_A^2$. Now in order to approximate this Gaussian pdf by pdf $a*Chi(ax,n)$, n should be expressed by $M_A$ and $\sigma_A$, consequently $n=2M_A/\sigma_A^2$, then n is the floating point value, leading to the possibility of using (AII. 3).

## APPENDIX III.

The efficiency of the MLBNR decision rule can be shown by comparing its performance with the Maximum Likelihood (ML) decision rule and with the traditional MLS decision. At first, calculate the error probability equation for the ML decision when the noise source state is known in the receiver. This means that the receiver decision is accomplished when it is known if the noise realization number $i - \xi_i$ was generated by a background noise with variance $\sigma_1^2(i)$ or by an impulse noise with variance $\sigma_2^2(i)$. For given length N, and right signal vector and mistaken vector, and difference vector **e**, an error occurs if the probability of the right vector to be transmitted is less than the mistaken one

$$P_{eML} = P\left[\prod_{i=1}^{N}\frac{e^{-\xi_i^2/(2\sigma(i)^2)}}{\sigma(i)\sqrt{2\pi}} < \prod_{i=1}^{N}\frac{e^{-(\xi_i+e_i)^2/(2\sigma(i)^2)}}{\sigma(i)\sqrt{2\pi}}\right], \quad \text{(AII. 1)}$$

by taking a logarithm, we have

$$P_{eML} = P\left[\sum_{i=1}^{N}\frac{\xi_i^2}{2\sigma^2(i)} > \sum_{i=1}^{N}\frac{(\xi_i+e_i)^2}{2\sigma^2(i)}\right]. \quad \text{(AII. 2)}$$

Now, by averaging for all possible noise sets, the final answer is

$$P_{eML} = \sum_{i_N=0}^{i_N=1}\ldots\sum_{i_1=0}^{i_1=1}[i_1(1-\varepsilon)+(1-i_1)\varepsilon]\ldots[i_N(1-\varepsilon)+(1-i_N)\varepsilon]*Q(\frac{1}{2}\sum_{j=1}^{N}\frac{e_j^2}{(\sigma_1 i_j + (1-i_j)\sigma_2)^2}). \quad \text{(AII. 3)}$$

For the MLBNR rule, only background noise realizations are relevant, since the reduced form of (AII. 3) gives error probability for MLBNR ( in order to increase the capability of the receiver to deal with the case when all noise realizations are formatted by impulse noise, the MLS detection for these realizations is introduced into the decision. In addition, note that it is implemented built-in in Robust procedures)

$$P_{eMLBNR} = \varepsilon^N Q\left(\frac{1}{2}\sqrt{\sum_{j=1}^{N}\frac{i_j e_j^2}{\sigma_1^2}}\right) + \sum_{i_{N-1}=0}^{i_{N-1}=1}\ldots\sum_{i_1=0}^{i_1=1}[i_1(1-\varepsilon)+(1-i_1)\varepsilon][i_{N-1}(1-\varepsilon)+(1-i_{N-1})\varepsilon]*Q\left(\frac{1}{2}\sqrt{\sum_{j=1}^{N}\frac{i_j e_j^2}{\sigma_1^2}}\right). \quad \text{(AII. 4)}$$

By taking into account that $\sigma_1 << \sigma_2$, which is received from the definition of impulse noise, we have an important conclusion from (AII. 4) and (AII. 3)

$$P_{eMLBR} \approx P_{eML}. \quad \text{(AII. 5)}$$

In order to show how far the MLBNR is from the MLS decision, it is necessary to calculate the probability of error for MLS

$$P_{MLS} = P\left[\sum_{i=1}^{N}\xi_i^2 > \sum_{i=1}^{N}(\xi_i + e_i)^2\right], \qquad (AII.\ 6)$$

which leads to

$$P_{eMLS} = P\left(\left(\xi, 0, \sum_{i=1}^{N}\sigma^2(i)e_i^2 > \frac{1}{2}\sum_{i=1}^{N}e_i^2\right)\right) = Q\left(\frac{\frac{1}{2}\sum_{i=1}^{N}e_i^2}{\sqrt{\sum_{i=1}^{N}\sigma^2(i)e_i^2}}\right)$$

$$(AII.\ 7)$$

For only background noise with variance $\sigma^2$, $P_{eMLS}$ reduced to the known probability of error in Gaussian noise

$$P_{eMLSB} = P\left[(\xi, 0, 1) > \frac{1}{2}\frac{\sqrt{\sum_{i=1}^{N}e_i^2}}{\sigma}\right] = Q\left(\frac{\frac{1}{2}\sqrt{\sum_{i=1}^{N}e_i^2}}{\sigma}\right). \qquad (AII.\ 8)$$

It is complicated to deduce the closed form for the ML case when the receiver does not know the kind of noise realization, i.e. if it was generated by background or impulse noise; in a general form this probability of mistake is

$$P^{u}_{eML} = P\left[\prod_{i=1}^{N}\rho(\xi_i) < \prod_{i=1}^{N}\rho(\xi_i + e_i)\right] \qquad (AII.\ 9)$$

It is possible to show the potential of the MLBNR decision by the next example of a simple vector communication system, when : 1, 1, 1 is transmitted for 1 and -1, -1, -1 for 0, and when noise realizations are undependable. In this case the probabilities of error are

$$P^{example}_{eML} = (1-\varepsilon)Q\left(\frac{\sqrt{3}}{\sigma_1}\right) + 3\varepsilon(1-\varepsilon)^2 Q\left(\sqrt{\frac{2}{\sigma_1^2}+\frac{1}{\sigma_2^2}}\right) + 3\varepsilon^2(1-\varepsilon)Q\left(\sqrt{\frac{1}{\sigma_1^2}+\frac{2}{\sigma_2^2}}\right) + \varepsilon^3 Q\left(\frac{\sqrt{3}}{\sigma_2}\right), \qquad (AII.\ 10)$$

$$P^{example}_{MLBNR} = (1-\varepsilon)^3 Q\left(\frac{\sqrt{3}}{\sigma_1}\right) + 3\varepsilon(1-\varepsilon)^2 Q\left(\frac{\sqrt{2}}{\sigma_1}\right) + 3\varepsilon^2(1-\varepsilon)Q\left(\frac{1}{\sigma_1}\right) + \varepsilon^3 Q\left(\frac{\sqrt{3}}{\sigma_2}\right), \qquad (AII.\ 11)$$

$$P^{example}_{eMLS} = (1-\varepsilon)^3 Q\left(\frac{\sqrt{3}}{\sigma_1}\right) + 3\varepsilon(1-\varepsilon)^2 Q\left(\frac{3}{\sqrt{2\sigma_1^2+\sigma_2^2}}\right) + 3\varepsilon^2(1-\varepsilon)Q\left(\frac{3}{\sqrt{\sigma_1^2+2\sigma_2^2}}\right) + \varepsilon^3 Q\left(\frac{\sqrt{3}}{\sigma_2}\right), \qquad (AII.\ 12)$$

$$P^{example}_{eMLSB} = Q\left(\frac{\sqrt{3}}{\sigma}\right), \qquad \text{(AIII. 13).}$$

From Fig. AIII. 1, it is possible to see the example of the quantitative analysis of performance relations between mentioned detectors.

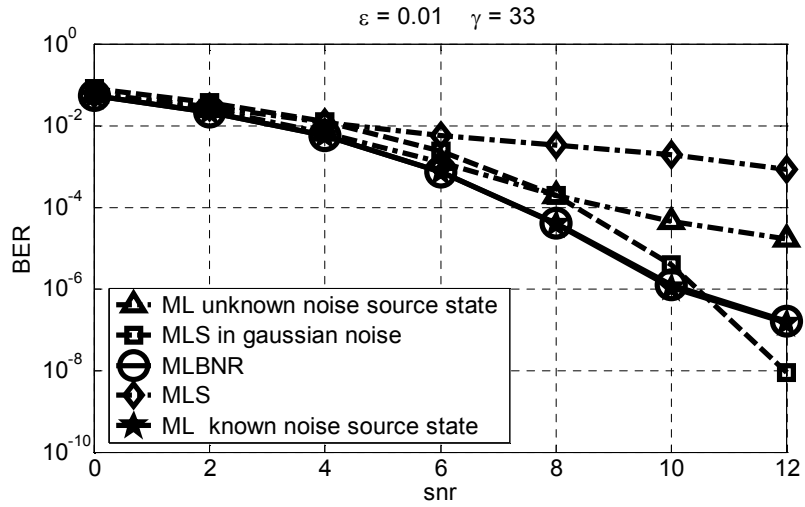

gh

Fig. AIII. 1

Fig. AIII. 1 Example of quantitative analysis of performance relations: the ML decoder when it is unknown if the noise sample was generated by background or impulse noise; for the ML decoder when it is known if the noise sample was generated by background or impulse noise; for the MLS decoder in impulsive noise and for the MLS decoder in only background noise.

The main results are:
- the MLBNR and ML are very close for known state noise source, which means that the MLBNR is as powerful as the most powerful rule for the Gaussian $\varepsilon$-tail, so, hopefully the MLBNR will be efficient for other tails, as efficient as it is for the investigated Laplas tail;
- Despite the fact the MLBNR detector ignores some information about the probability of the transmitted signal, the MLBNR receiver is better than the ML receiver for unknown state noise source, which means that it is possible to expect much better results from Robust algorithms based on this rule than from the MLS detector;
- the MLBNR receiver is much better than the MLS receiver ;
- Sometimes, the ML and MLBNR decisions in impulsive noise provide a better error rate than MLS in Gaussian noise. It therefore shows that it worthwhile investigating this issue of design decoding algorithms based on taking into account the non-Gaussianity of noise, since the achievable performance can be better than in Gaussian noise .

## VI. CONCLUDING REMARKS

A Robust method for signal decoding in a mixture of Gaussian background and impulse noises was proposed. Simulations and analytical analysis show that Robust in a statistical sense approach, based on a Maximum Likelihood of background noise residuals, offers significant performance gain, when the loss is small in only background noise.

An approach to an analytical performance analysis of MLS MIMO schemes in Gaussian noise and $\varepsilon$-noise was also developed.

The example of suggested methods for using a 2Tx2Rx Alamouti's scheme has been considered.

## VII. ACKNOWLEDGMENT

The author is grateful to Mr. G. Levin, of the University of Ottawa, for encouraging him to do this work and for our useful discussions. Thanks also to Massimo Allegritti, of the Politecnico di Milano, for the use of his Matlab code simulation. The code was taken from the Matlab Central site as the starting point in building these investigation simulations.